\documentclass[trackchanges]{aastex701}

\usepackage{amsmath} 
\usepackage{CJKutf8}

\begin{document}

\title{Revisiting the Radial Velocities of Nearby Open Clusters using {\it Gaia} DR3}

\author[0000-0003-1864-8721]{Tong Tang(\begin{CJK*}{UTF8}{gbsn}唐通\end{CJK*})}
\affiliation{XinJiang Astronomical Observatory, Chinese Academy of Sciences,150 Science 1-Street, Urumqi, Xinjiang 830011, PR China}
\affiliation{School of Astronomy and Space Science, University of Chinese Academy of Sciences, No. 19A, Yuquan Road, Beijing 100049, PR China}
\email{tangtong@xao.ac.cn}

\author[0000-0001-7134-2874]{Yu Zhang(\begin{CJK*}{UTF8}{gbsn}张余\end{CJK*})}
\altaffiliation{Corresponding authors: zhy@xao.ac.cn, jzhong@shao.ac.cn}
\affiliation{XinJiang Astronomical Observatory, Chinese Academy of Sciences,150 Science 1-Street, Urumqi, Xinjiang 830011, PR China}
\affiliation{School of Astronomy and Space Science, University of Chinese Academy of Sciences, No. 19A, Yuquan Road, Beijing 100049, PR China}
\email{zhy@xao.ac.cn}

\author[0000-0001-5245-0335]{Jing Zhong(\begin{CJK*}{UTF8}{gbsn}钟靖\end{CJK*})}
\altaffiliation{Corresponding authors: zhy@xao.ac.cn, jzhong@shao.ac.cn}
\affiliation{Astrophysics Division, Shanghai Astronomical Observatory, Chinese Academy of Sciences, 80 Nandan Road, Shanghai 200030, PR China}
\affiliation{School of Astronomy and Space Science, University of Chinese Academy of Sciences, No. 19A, Yuquan Road, Beijing 100049, PR China}
\email{jzhong@shao.ac.cn}

\author[0009-0002-4686-8115]{Guimei Liu(\begin{CJK*}{UTF8}{gbsn}刘桂梅\end{CJK*})}
\affiliation{XinJiang Astronomical Observatory, Chinese Academy of Sciences,150 Science 1-Street, Urumqi, Xinjiang 830011, PR China}
\affiliation{School of Astronomy and Space Science, University of Chinese Academy of Sciences, No. 19A, Yuquan Road, Beijing 100049, PR China}
\affiliation{Department of Astrophysics, University of Vienna, T\"urkenschanzstra{\ss}e 17, 1180 Wien, Austria}
\email{liuguimei20@mails.ucas.ac.cn}

\author[0000-0003-3713-2640]{Songmei Qin(\begin{CJK*}{UTF8}{gbsn}秦松梅\end{CJK*})}
\affiliation{Astrophysics Division, Shanghai Astronomical Observatory, Chinese Academy of Sciences, 80 Nandan Road, Shanghai 200030, PR China}
\affiliation{School of Astronomy and Space Science, University of Chinese Academy of Sciences, No. 19A, Yuquan Road, Beijing 100049, PR China}
\email{qinsongmei@shao.ac.cn}

\author[0009-0007-9332-5865]{Shusong Wang(\begin{CJK*}{UTF8}{gbsn}王述松\end{CJK*})}
\affiliation{XinJiang Astronomical Observatory, Chinese Academy of Sciences,150 Science 1-Street, Urumqi, Xinjiang 830011, PR China}
\affiliation{School of Astronomy and Space Science, University of Chinese Academy of Sciences, No. 19A, Yuquan Road, Beijing 100049, PR China}
\email{Corvo-Wang@outlook.com}

\author[0009-0007-9332-5865]{Li Chen(\begin{CJK*}{UTF8}{gbsn}陈力\end{CJK*}) }
\affiliation{Astrophysics Division, Shanghai Astronomical Observatory, Chinese Academy of Sciences, 80 Nandan Road, Shanghai 200030, PR China}
\affiliation{School of Astronomy and Space Science, University of Chinese Academy of Sciences, No. 19A, Yuquan Road, Beijing 100049, PR China}
\email{chenli@shao.ac.cn}

\begin{abstract}

Open clusters (OCs) are essential laboratories for probing stellar dynamics and tracing the structure and evolution of the Milky Way. Accurate measurements of their average radial velocities (RVs) and RV dispersions are crucial for estimating dynamical masses, orbital evolution, and overall kinematic states. {\it Gaia} Data Release 3 (DR3) provides an unprecedented volume of high-precision RVs. However, when applied to OCs, {\it Gaia} DR3 RVs often yield unusually large and overestimated RV dispersions. This inflation is primarily driven by RV measurement systematics for hot and faint stars, as well as unresolved binary contamination. To mitigate this, we revisit the average RVs and RV dispersions of OCs within 500\,pc of the Sun using {\it Gaia} DR3. We evaluate the reliability of RV measurements and employ a color-based filtering method. By selecting member stars within an intrinsic color range of $0.2 \le (BP-RP)_{0} \le 1.2$\,mag, we exclude hot and cool stars with less reliable RVs. This filtering significantly reduces the inferred RV dispersions while maintaining average RVs consistent with previous literature. Specifically, the median RV dispersion decreases by 26\%, dropping from 3.76\,km\,s$^{-1}$ prior to filtering to 2.79\,km\,s$^{-1}$ afterward. These filtered RV dispersions remain systematically larger than tangential velocity dispersions, likely due to residual biases and undetected binaries. However, RV dispersions measured exclusively from red clump giants (found in 6 clusters) are remarkably small ($\lesssim 1.6$\,km\,s$^{-1}$) and align closely with tangential dispersions ($\lesssim 1$\,km\,s$^{-1}$). Ultimately, we provide a practical, {\it Gaia}-only strategy to derive more realistic RV dispersions for OCs, identifying red clump giants as exceptionally high-fidelity kinematic tracers for robust cluster studies.

\end{abstract}

\keywords{\uat{Open star clusters}{1160} --- \uat{Radial velocity}{1332} --- \uat{Stellar kinematics}{1608} --- \uat{Astronomy data analysis}{1858}}

\defcitealias{2023ApJS..265...12Q}{Qin23}

\section{Introduction}

Open clusters (OCs) serve as crucial tracers of the structure and evolution of the Milky Way. Their average radial velocities (RVs) are key parameters for determining full six-dimensional phase-space coordinates, providing the foundation for reconstructing OC kinematics within the Galactic potential \citep{2018A&A...619A.155S,2021A&A...647A..19T,2022A&A...658A..14C}. Furthermore, the RVs of member stars are instrumental in identifying kinematic signatures such as rotation, expansion, or contraction, thereby offering critical observational constraints for theories of cluster formation and evolution \citep{2019MNRAS.483.2197K,2024A&A...687A..89J}. Moreover, the RV dispersion within OCs provides a key tracer of their internal dynamics, facilitating the estimation of the cluster dynamical masses and overall dynamical states for clusters \citep{2018A&A...615A..37B,2024AJ....167..212K}. Consequently, obtaining precise measurements of the average RVs and RV dispersions of OCs is of paramount importance; these metrics not only yield robust evaluations of dynamical masses, orbital evolution, and kinematic states of clusters, but also impose vital observational constraints on the structure, rotation curve, and evolutionary history of the Milky Way.

The ambitious European Space Agency (ESA) mission {\it Gaia} \citep{2016A&A...595A...1G} has provided unprecedented insights into the physical properties of Milky Way stars. Based on 34 months of satellite operations, {\it Gaia} Data Release 3 (DR3, \citealt{2023A&A...674A...1G}) provides astrometry and broadband photometry for 1.8 billion objects, additionally reporting RVs for 33 million stars with $G_{\rm RVS}$ $<$ 14\,mag and 3100\,K $<$ $T_{\rm eff}$ $<$ 14500\,K. The wavelength range of the {\it Radial Velocity Spectrometer} (RVS) spectra is 846-870\,nm with a resolution of $R$ = $\lambda/\Delta\lambda\sim$11500. Stellar RVs are derived through cross-correlation techniques utilizing all available spectral lines within this range, with the signal being heavily dominated by the Ca\,\rm\uppercase\expandafter{\romannumeral2} triplet \citep{2018A&A...616A...6S,2018A&A...616A...5C}. These velocities achieve a median formal precision of 1.3\,km $\rm s^{-1}$ at $G_{\rm RVS}$ = 12\,mag and 6.4\,km $\rm s^{-1}$ at $G_{\rm RVS}$ = 14\,mag \citep{2023A&A...674A...5K}. Although {\it Gaia} DR3 provides an unparalleled number of stellar RVs, the RV precision decreases at the faint end due to scattered light contamination \citep{2023A&A...674A...5K}, and systematic biases occur for hot stars because the broad and shallow lines of the Paschen series affect the RV determination \citep{2023A&A...674A...7B}.

The integration of {\it Gaia} era high-precision astrometric and photometric data with clustering algorithms \citep{2014A&A...561A..57K,2017JOSS....2..205M,2021A&A...650A.109P,2021A&A...646A.104H} has facilitated the discovery of an ever-growing number of OCs \citep{2018A&A...618A..93C,2020A&A...640A...1C,2019ApJS..245...32L,2019JKAS...52..145S,2020A&A...635A..45C,2022A&A...661A.118C,2021MNRAS.504..356D,2021RAA....21...45Q,2023ApJS..265...12Q,2022ApJS..260....8H,2022ApJS..262....7H,2023A&A...673A.114H}. In a recent study, \citet{2023A&A...673A.114H} compiled homogeneous catalog containing over 7000 clusters based on {\it Gaia} DR3 data, designating 3530 OCs as a high-quality sample. Among these, 1559 OCs possess RV measurements for more than five member stars. However, as revealed by this extensive dataset, the RV dispersions for OCs remain systematically overestimated due to an issue inherent to the {\it Gaia} DR3 data. Previous studies relying on high-resolution spectroscopic RV measurements of member stars have demonstrated that, even for certain young OCs, the 3D velocity dispersion ranges from 0.6 to 7.4\,km\,s$^{-1}$ \citep{2018A&A...615A..37B,2024MNRAS.533..705W}. The RV dispersions derived from {\it Gaia} DR3 for these OCs are substantially larger than expected, likely due to a combination of systematic RV measurement errors and contamination from unresolved binaries \citep{2012A&A...547A..35C,2023A&A...674A..32B}. The influence of measurement systematics on the cluster RV dispersion primarily originates from biases inherent in {\it Gaia} observations of hot and faint stars. Furthermore, the observed RVs of unresolved binaries contain an additional orbital component that artificially inflates the inferred cluster RV dispersion, an effect strongly contingent upon the binary fraction \citep{2008AJ....135.2264G,2024AJ....167..212K}.

It is widely recognized that utilizing {\it Gaia} DR3 RVs for OCs results in significantly inflated estimates of velocity dispersions. In this work, we re-examine the average RVs and RV dispersions of OCs within 500\,pc of the solar neighborhood using {\it Gaia} DR3 data. We employ a color-based filter to retain only members with reliable RV measurements, thereby markedly suppressing the extra dispersion caused by measurement errors. Furthermore, we investigate the impact of binaries on the average RV and RV dispersion of these clusters. Additionally, we find that the RV dispersion derived from red clump giants is in excellent agreement with the tangential velocity dispersion.

The paper is structured as follows. In Sect.~\ref{sec:Sqc}, we describe the sample used and quality control in this work. In Sect.~\ref{sec:AROC}, we present the results for the average RVs and RV dispersions of OCs and discuss them in detail. In Sect.~\ref{sec:CD}, we compared our results with those of previous studies and discussed the binary star sample and the red clump giant sample. Finally, we summarize our findings in Sect.~\ref{sec:C}.

\section{Sample and quality control}\label{sec:Sqc}

Based on the {\it Gaia} DR3 data, \citet[hereafter \citetalias{2023ApJS..265...12Q}]{2023ApJS..265...12Q} utilized the pyUPMASK (Unsupervised Photometric Membership Assignment in Stellar Clusters, ~\citealt{2014A&A...561A..57K,2021A&A...650A.109P}) and HDBSCAN (Hierarchical Density-Based Spatial Clustering of Applications with Noise, ~\citealt{2017JOSS....2..205M}) clustering algorithms to systematically search for OCs at Galactic latitude of $|b| < 30^\circ$ within 500\,pc of the Sun. They presented a catalog of 324 OCs, including 101 newly discovered clusters. This catalog provides cluster ages, distance modulus, and reddening values derived via visual isochrone fitting. This is a relatively complete sample of nearby star clusters at present. The OC sample analyzed in our study is drawn from this catalog. 

For this sample, we performed a visual inspection of the color-magnitude diagrams (CMDs) utilizing the age parameters provided by \citetalias{2023ApJS..265...12Q}, and excluded any clusters exhibiting evident mismatches. Using the distance modulus $(m-M)_0$ and redding $E(B-V)$, we derive the extinction in {\it G} band as  $A_{\rm G}=2.74 \times E(B-V)$ and the $G_{\rm BP}-G_{\rm RP}$ redding $E(G_{\rm BP}-G_{\rm RP})=1.339 \times E(B-V)$ \citep{2018MNRAS.479L.102C,2019A&A...624A..34Z}. For each member star, the absolute magnitude and intrinsic color are subsequently calculated using the following equations:
\begin{equation}
\begin{split}
M_{\rm G} &= G_{\rm obs} - (m-M)_0 - A_{\rm G}, \\
(BP-RP)_{0} &= (BP-RP)_{\rm obs} - E(BP-RP).
\end{split}
\end{equation}

\begin{figure*}[!htbp]
\centering
\includegraphics[width=0.6\textwidth]{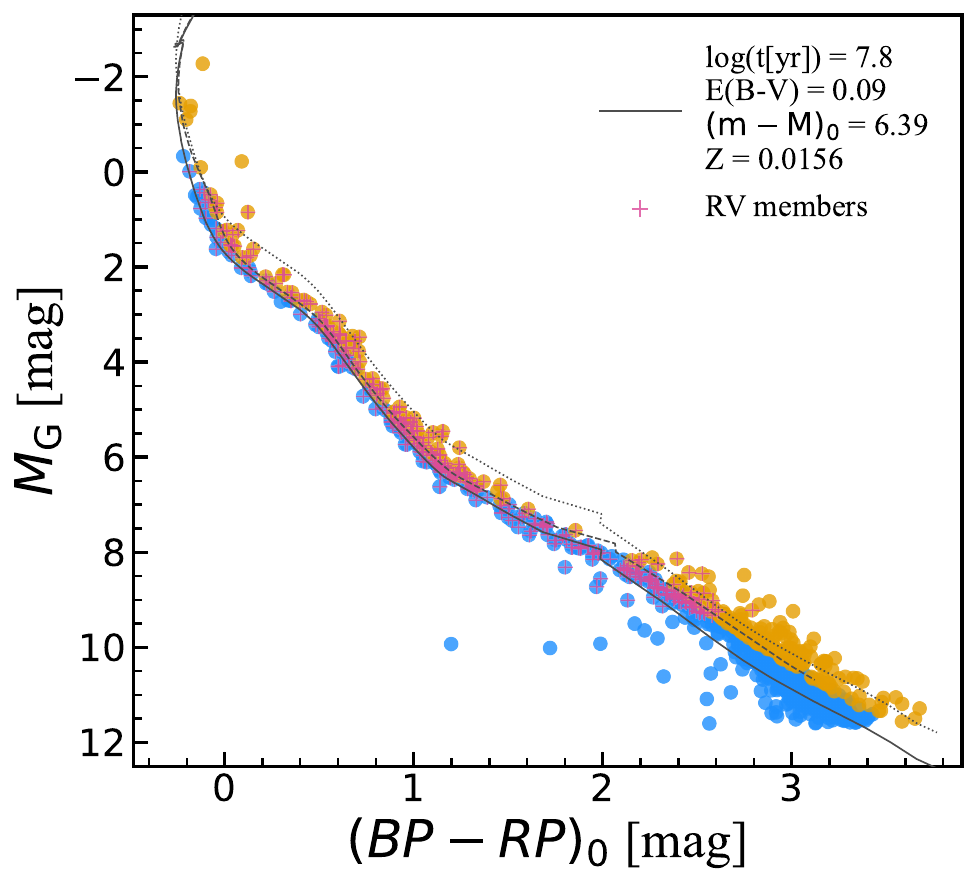}
\caption{The color-absolute magnitude diagram of members of OCSN 259 (Roslund 6). The solid, dashed, and dotted lines show the best-fit isochrone, the q=0.5 binary sequence, and the q=1 binary sequence, respectively, where q is the mass ratio of the secondary to primary component. The blue and orange dots are the binary members with mass ratios less than and larger than 0.5. The pink cross represents that this star has an RV measurement provided by {\it Gaia} DR3.}
\label{Fig1}
\end{figure*}

Subsequently, we retrieved the PARSEC 1.2S isochrones \citep{2017ApJ...835...77M} with the {\it Gaia} photometric system \citep{2021yCat..36490003R} from the  CMD\,3.9\footnote{\url{http://stev.oapd.inaf.it/cmd}}. Fig.~\ref{Fig1} illustrates a representative cluster selected in this study. To ensure a robust statistical analysis, we restricted our sample to clusters containing a minimum of 20 member stars with available RV measurements. Ultimately, we retained a final sample of 246 OCs comprising 53584 member stars, 20383 of which possess {\it Gaia} DR3 RV measurements.

\section{Analysis of the RVs of open clusters.}\label{sec:AROC}

The stellar RVs provided by {\it Gaia} DR3 are derived from medium-resolution spectra acquired by the RVS, yielding an unprecedented sample of over 33 million measurements. The median formal RV precision of these RVs is 1.3\,km $\rm s^{-1}$ at $G_{\rm RVS}$=12\,mag and 6.4\,km $\rm s^{-1}$ at $G_{\rm RVS}$=14\,mag \citep{2023A&A...674A...5K}. Recent studies indicate that {\it Gaia} DR3 RVs suffer from substantial uncertainties for hot and faint stars \citep{2023A&A...674A...5K,2023A&A...674A...7B}. When analyzing OCs, these observational uncertainties will lead to an increase in the inferred cluster RV dispersions. To mitigate this, we apply a color-based filter—serving as a proxy for effective temperature—to the member stars before calculating the average RVs and RV dispersions of the OCs.

\begin{figure*}[!htbp]
\centering
\includegraphics[width=0.7\textwidth]{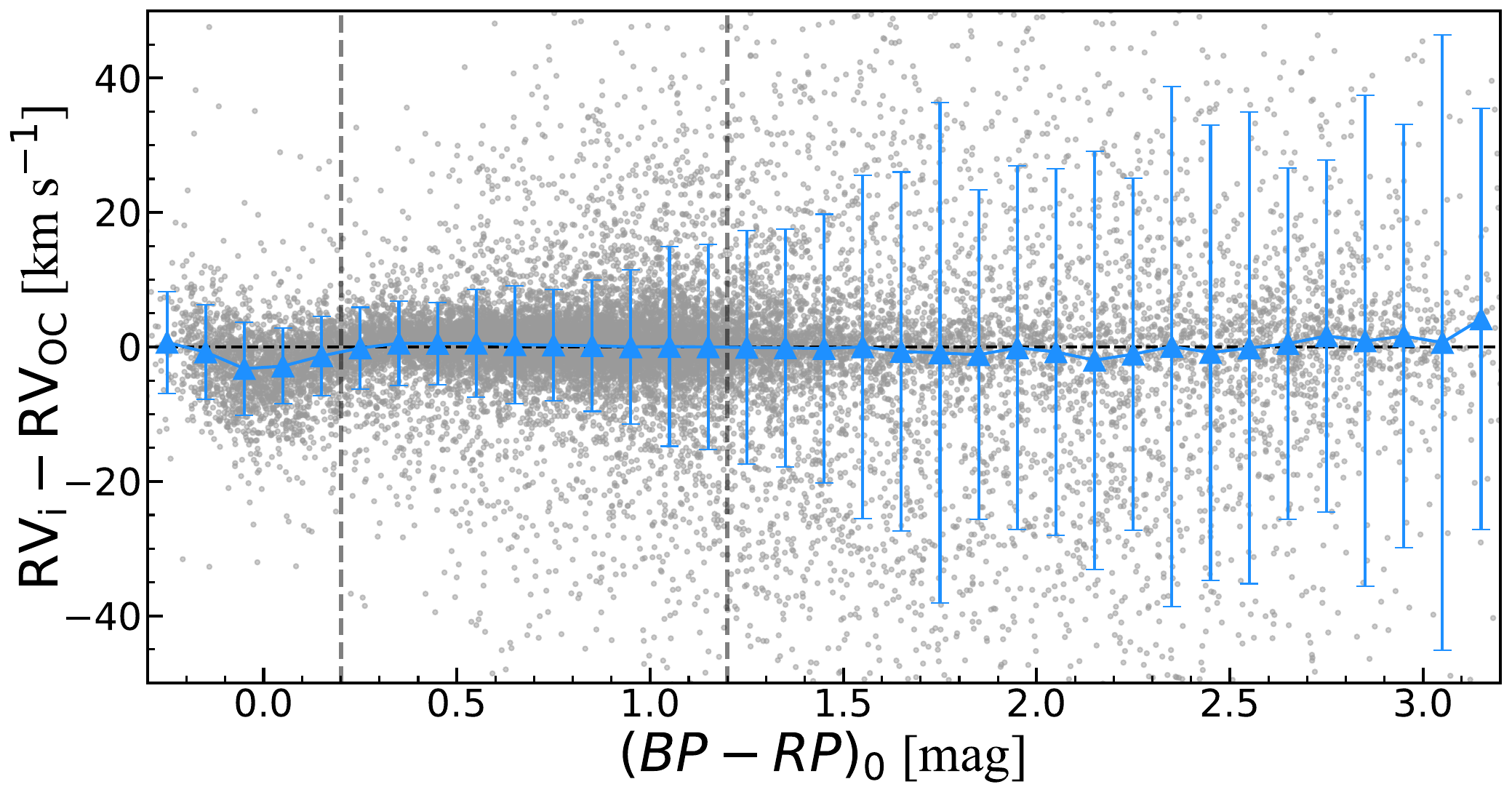}
\caption{RV offsets of member stars (RV$_{\rm i}$) relative to the cluster median RV (RV$_{\rm OC}$) as a function of intrinsic color. Blue triangles with error bars show the median and the Root Mean Square Deviation (RMSD) around the median of the offsets with a step of 0.1\,mag in the intrinsic color range of [-0.3, 3.2]\,mag. The two vertical dashed lines correspond to $(BP-RP)_0$ of 0.2 and 1.2\,mag, respectively.}
\label{Fig2}
\end{figure*}

To analyze the offsets and dispersions of member star RVs relative to the cluster average, we define the median RV of an OC ($\rm RV_{\rm OC}$) as the median RVs of member stars. Fig.~\ref{Fig2} illustrates the distribution of RV residuals (the difference between members' star RVs and the $\rm RV_{\rm OC}$ of their respective) as a function of intrinsic color. The data clearly reveal that hot member stars exhibit a systematic offset of up to approximately 3\,km\,s$^{-1}$ near $(BP-RP)_0=0$\,mag. Conversely, the systematic RV offset approaches zero, around $(BP-RP)_0=0.2$\,mag. Therefore, we adopt $(BP-RP)_0=0.2$\,mag as the boundary on the blue end. For cool member stars, the scatter of RVs is particularly significant, and the number of such stars is relatively small. This is likely due to the difficulty in measuring their absorption lines, as other molecular lines influence the absorption lines in the RVS spectra of cool stars \citep{2018A&A...616A...5C}. Thus, we adopt $(BP-RP)_0=1.2$\,mag as the boundary on the red end. 

\begin{figure*}[!htbp]
\centering
\includegraphics[width=0.9\textwidth]{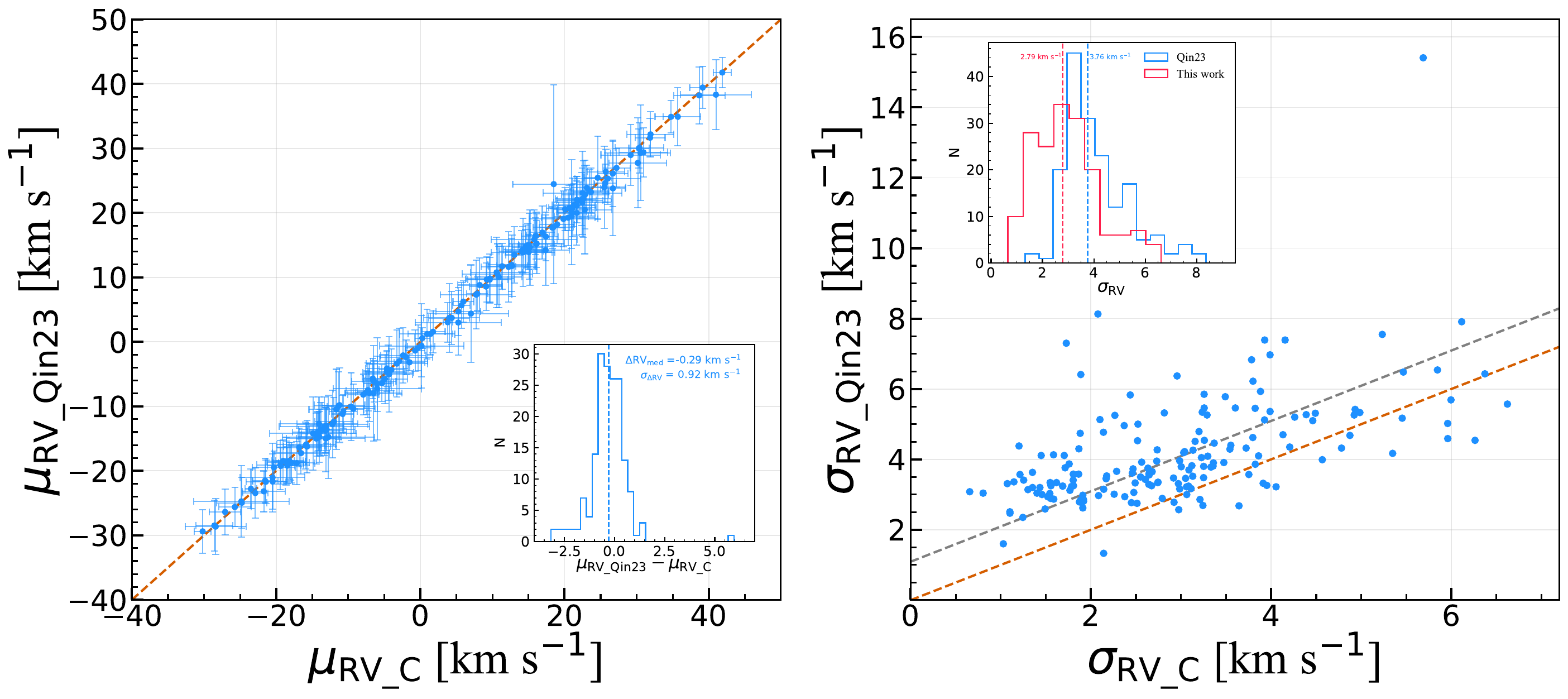}
\caption{Comparison of average RVs (left panel) and RV dispersions (right panel) of cluster after color-based filtering of member stars to results of \citetalias{2023ApJS..265...12Q}. Distributions of the average RV differences are shown as inset histograms in the left panel. The inset histogram of the RV dispersion is shown in the right panel. The red dashed lines are 1:1 scale lines, while the gray dashed line indicates the same relation shifted upward by the offset of approximately 1.1\,km\,s$^{-1}$.}
\label{Fig3}
\end{figure*}

\begin{deluxetable*}{ccccccccccrrc}
\tabletypesize{\footnotesize} 
\tablecaption{The average RVs and RV dispersions of the 171 OCs derived from color-based filtering of member stars. \label{Table1}}
\renewcommand{\arraystretch}{1.1}
\tablehead{
\colhead{Name} & \colhead{RAdeg} & \colhead{DEdeg} & \colhead{$\varpi$} & \colhead{e\_$\varpi$} & \colhead{log(t)} & \colhead{$E(B-V)$} & \colhead{$(m-M)_0$} & \colhead{$\sigma_{\rm R.A.}$} & \colhead{$\sigma_{\rm Decl.}$} & \colhead{$\mu_{\rm RV\_C}$} & \colhead{$\sigma_{\rm RV\_C}$} & \colhead{N\_RV} \\
\colhead{} & \colhead{[deg]} & \colhead{[deg]} & \colhead{[mas]} & \colhead{[mas]} & \colhead{[log(yr)]} & \colhead{[mag]} & \colhead{[mag]} & \colhead{[km\,s$^{-1}$]} & \colhead{[km\,s$^{-1}$]} & \colhead{[km\,s$^{-1}$]} & \colhead{[km\,s$^{-1}$]} & \colhead{} \\
\colhead{(1)} & \colhead{(2)} & \colhead{(3)} & \colhead{(4)} & \colhead{(5)} & \colhead{(6)} & \colhead{(7)} & \colhead{(8)} & \colhead{(9)} & \colhead{(10)} & \colhead{(11)} & \colhead{(12)} & \colhead{(13)}
}
\startdata
OCSN 10  & 275.41 & 14.82 & 2.73 & 0.16 & 7.45 & 0.19 & 8.49 & 0.64 & 0.79 & -13.11 & 5.45 & 22 \\
OCSN 12  & 276.54 & 21.08 & 2.31 & 0.16 & 7.45 & 0.17 & 8.71 & 0.95 & 1.29 & -16.55 & 3.95 & 35 \\
OCSN 13  & 291.75 & 21.1  & 2    & 0.07 & 6.9  & 0.15 & 8.87 & 0.82 & 0.96 & -9.29  & 3    & 21 \\
OCSN 15  & 284.96 & 27.34 & 3.88 & 0.24 & 8.1  & 0.2  & 7.46 & 0.66 & 1.46 & -25.72 & 2.26 & 23 \\
OCSN 16  & 283.56 & 32.18 & 2.68 & 0.12 & 7.45 & 0.11 & 8.18 & 0.66 & 0.56 & -15.91 & 2.68 & 35 \\
\dots    & \dots  & \dots & \dots & \dots & \dots & \dots & \dots & \dots & \dots & \dots & \dots & \dots \\
\enddata
\tablecomments{Columns (1)-(8) list the information on the projected position, parallax, age, reddening, and distance modulus of the OCs from \citetalias{2023ApJS..265...12Q}. Columns (9)-(10) present our estimates of the tangential velocity dispersion. Columns (11)-(13) present the average RV, the standard deviation of RV, and the number of stars used, respectively. The complete list of these clusters is available at the CDS.}
\end{deluxetable*}

To minimize the impact of RV measurement biases on the inferred cluster RV dispersions, we apply a color-based filtering scheme to exclude both hot and cool stars. Based on the distribution shown in Fig.~\ref{Fig2}, we retained member stars within the intrinsic color range of [0.2, 1.2]\,mag. 
Finally, we exclude OCs possessing fewer than 20 RV-measured member stars within this specified color regime, resulting in the retention of 171 clusters and the removal of 75.

To further demonstrate that our color-based filtering method effectively reduces the inferred RV dispersions of clusters, we compare our results with those of \citetalias{2023ApJS..265...12Q}. We performed the same Gaussian fitting procedure as \citetalias{2023ApJS..265...12Q} to the RVs of the color-based filtered member stars, which incorporates a $3\sigma$ clipping to reject velocity outliers progressively. By fitting the remaining stars, we derived the mean and standard deviation to represent the cluster average RV ($\mu_{RV\_C}$) and RV dispersion ($\sigma_{\rm RV\_C}$), respectively. 
Fig.~\ref{Fig3} illustrates the comparison of our results with those reported by \citetalias{2023ApJS..265...12Q}. As shown in the left panel of Fig.~\ref{Fig3}, our results for the average RVs of OCs are consistent with those of \citetalias{2023ApJS..265...12Q}, with a median offset of -0.29\,km\,s$^{-1}$ and a scatter of 0.92\,km\,s$^{-1}$. Conversely, the right panel of Fig.~\ref{Fig3} reveals that our derived RV dispersions are systematically lower than those reported by \citetalias{2023ApJS..265...12Q}, showing a median reduction of 1\,km\,s$^{-1}$. This systematic reduction further verifies the effectiveness of adopting color-based filtering for stellar R samples.

Although the comparison with \citetalias{2023ApJS..265...12Q} demonstrates the effectiveness of the color-based filtering, it is also informative to examine the resulting RV dispersion in the context of other kinematic indicators. We therefore compare the RV dispersion with the tangential velocity dispersion. However, the observed proper motions of OC member stars are also biased by the systemic center-of-mass motion and the perspective expansion effect \citep{2009A&A...497..209V,2019ApJ...870...32K}. To eliminate this observational bias, kinematic corrections are typically performed using the first-order approximation equations proposed by \citet{2009A&A...497..209V}. The first-order approximation equations for the proper motions of a member star are
\begin{equation}
\begin{split}
\Delta \mu_{\alpha*,i} &\approx \Delta \alpha_i \left( \mu_{\delta,0} \sin \delta_0 - \frac{V_{\mathrm{rad},0} \varpi_0}{\kappa} \cos \delta_0 \right), \\
\Delta \mu_{\delta,i} &\approx -\Delta \alpha_i \mu_{\alpha*,0} \sin \delta_0 - \Delta \delta_i \frac{V_{\mathrm{rad},0} \varpi_0}{\kappa}.
\end{split}
\end{equation}
where $\Delta \alpha_i = \alpha_i - \alpha_0$ and $\Delta \delta_i=\delta_i - \delta_0$, the $\alpha_0$, $\delta_0$, $\mu_{\alpha*,0}$, $\mu_{\delta_0}$, and $\varpi_0$ are the astrometric parameters of the cluster center, and $V_{\mathrm{rad},0}$ is the average RV of the cluster obtained from our fitting, and $\kappa=4.74$ represents the transformation factor required to convert a proper motion of 1 mas yr$^{-1}$ at a distance of 1 kpc to a transverse velocity of 1 km s$^{-1}$.

\begin{figure}[!htbp]
\centering
\includegraphics[width=0.95\textwidth]{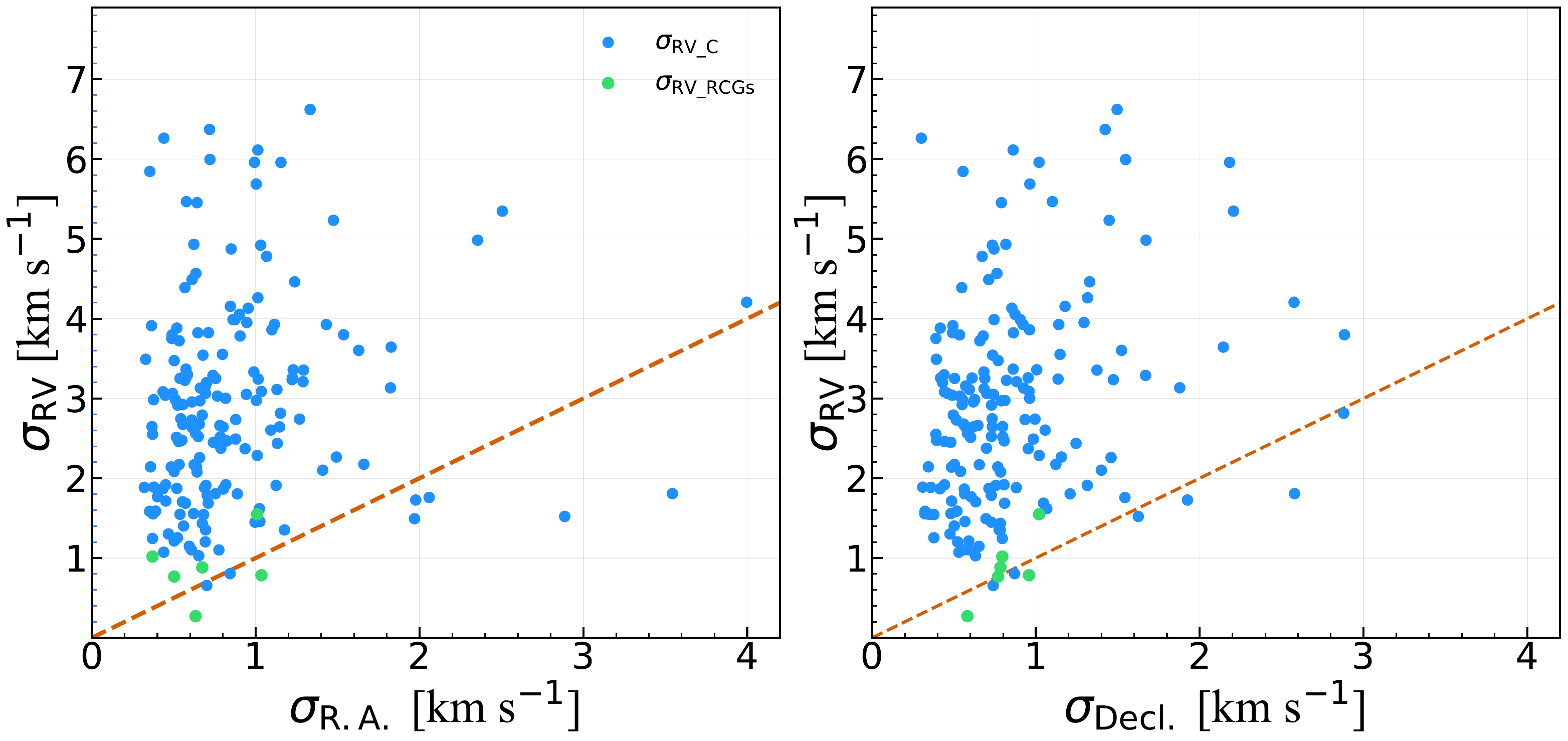}
\caption{Comparison of RV dispersions and tangential velocity dispersions. The blue points denote dispersions computed from color-based filtering of member stars, and the green points denote those derived from red clump giants. The red dashed lines are 1:1 scale lines.}
\label{Fig5}
\end{figure}

To evaluate the internal kinematics, the tangential velocity was estimated by converting the corrected proper motion components using the relation $4.74 \times \frac{\mu_{\alpha_i(\delta_i)}}{\varpi_{0}}$, where $\varpi_{0}$ represents the mean parallax of the cluster, and $\mu_{\alpha_i(\delta_i)}$ denotes the proper motion of member stars. These results are summarized in Table~\ref{Table1}. Our analysis reveals that the RV dispersions, even after rigorous color-based filtering, remain systematically larger than the tangential components (as illustrated by the blue dots in Fig.~\ref{Fig5}). Specifically, the sample exhibits a median of approximately 0.71\,km\,s$^{-1}$, whereas the RV dispersion is substantially larger. 
Physically, assuming kinematic isotropy for a dynamically relaxed stellar system, the intrinsic velocity dispersions in the radial and tangential directions are expected to be comparable \citep{2008gady.book.....B}. Observational equivalence, however, cannot be assumed. While astrometric proper motions depend primarily on geometric baselines and yield homogeneous errors, spectroscopic RV uncertainties are deeply coupled with intrinsic stellar properties (such as effective temperature and surface gravity). Variations in these parameters induce diverse spectral line broadening, leading to significantly larger RV errors across a vast stellar sample \citep{2018A&A...616A...5C,2023A&A...674A...5K}.
Consequently, this discrepancy suggests that the derived RV dispersions may not purely reflect the intrinsic dynamical state of the OCs. Such an inflation is likely dominated by the formal uncertainties in {\it Gaia} DR3 RVs for these member stars, coupled with the pervasive influence of undetected binaries.

In summary, our investigation demonstrates that the implementation of a color-based filtering scheme ($0.2 < (BP-RP)_0 < 1.2$\,mag) mitigates the systematic biases inherent in {\it Gaia} DR3 RVs, particularly for hot and cool member stars. Although this method yields a significant reduction of approximately 26\% in the median RV dispersion compared to previous catalogs, the resulting values remain systematically higher than the tangential velocity dispersions (a median of 0.71\,km\,s$^{-1}$). This significant discrepancy indicates that the observed RV dispersion is still dominated by extrinsic factors—primarily the formal measurement uncertainties of {\it Gaia} RVS spectra and the additional velocity components arising from the orbital motion of undetected binaries—rather than the OCs' intrinsic virial motion. Consequently, while {\it Gaia} DR3 provides an invaluable resource for average cluster velocities, we emphasize that its RVs should be treated with extreme caution in studies focusing on the internal dynamical states of OCs, as they likely represent an upper limit rather than a direct measurement of the true velocity dispersion.

\section{Comparison and discussion}\label{sec:CD}
\subsection{Comparison of clusters RV with other literature}

Driven by the growing availability of high-resolution spectroscopic survey data, numerous studies have investigated the RVs of OCs. Consequently, extensive catalogs of OC RVs have been compiled utilizing both space- and ground-based telescope observations \citep[e.g.,][]{2002A&A...389..871D,2005A&A...438.1163K,2013A&A...558A..53K,2018A&A...619A.155S,2020A&A...640A.127Z,2020AJ....159..199D,2021A&A...647A..19T,2022A&A...668A...4F,2022MNRAS.509.1664J,2022A&A...658A..14C}. To assess the reliability of the average RVs of OCs derived from the color-based filtering, we compared them with RV reported in other studies. Specifically, we benchmark our findings against the extensive RV catalog compiled by \citet{2021A&A...647A..19T}, which integrates {\it Gaia} measurements with ground-based spectroscopic surveys, including the {\it Gaia}-ESO (GSE, \citealt{2013Msngr.154...47R}), the Apache Point Observatory Galactic Evolution Experiment survey (APOGEE, \citealt{2020ApJS..249....3A}), the Radial Velocity Experiment (RAVE, \citealt{2017AJ....153...75K}), the GALactic Archaeology with HERMES (GALAH, \citealt{2021MNRAS.506..150B}). Adopting the cluster memberships from \citet{2020A&A...640A...1C}, they systematically evaluated and corrected the zero-point offsets across different RV data sources, compiling what was then the most comprehensive RV catalog for OCs. They provided weighted mean RVs for 1382 OCs, 38\% with a highly reliable RV based on more than three stars, and constructed a high precision 6D kinematic sample based on these RVs. Additionally, we compare our findings with the OC RV catalogs derived from the Large Sky Area Multi-Object Fiber Spectroscopic Telescope Low-Resolution Spectroscopic survey (LAMOST LRS, \citealt{2012RAA....12.1197C,2012RAA....12..723Z}) by \citet{2020A&A...640A.127Z} and \citet{2022A&A...668A...4F}. These studies determined the average RV for 295 and 386 OCs from the low-resolution LAMOST LRS DR5 and DR8 datasets, respectively. It is worth noting that 226 OCs are common to both works, with a median RV offset of just 0.06\,km\,$^{-1}$. Furthermore, based on the spectral parameters from LAMOST Medium-Resolution Spectroscopic Survey (MRS) DR11, \cite{2026RAA....26e5001Z} constructed a radial velocity catalog of 1033 OCs, where 50\% of the clusters have three or more member stars with radial velocity measurements.

\begin{figure*}[!htbp]
\centering
\includegraphics[width=0.99\textwidth]{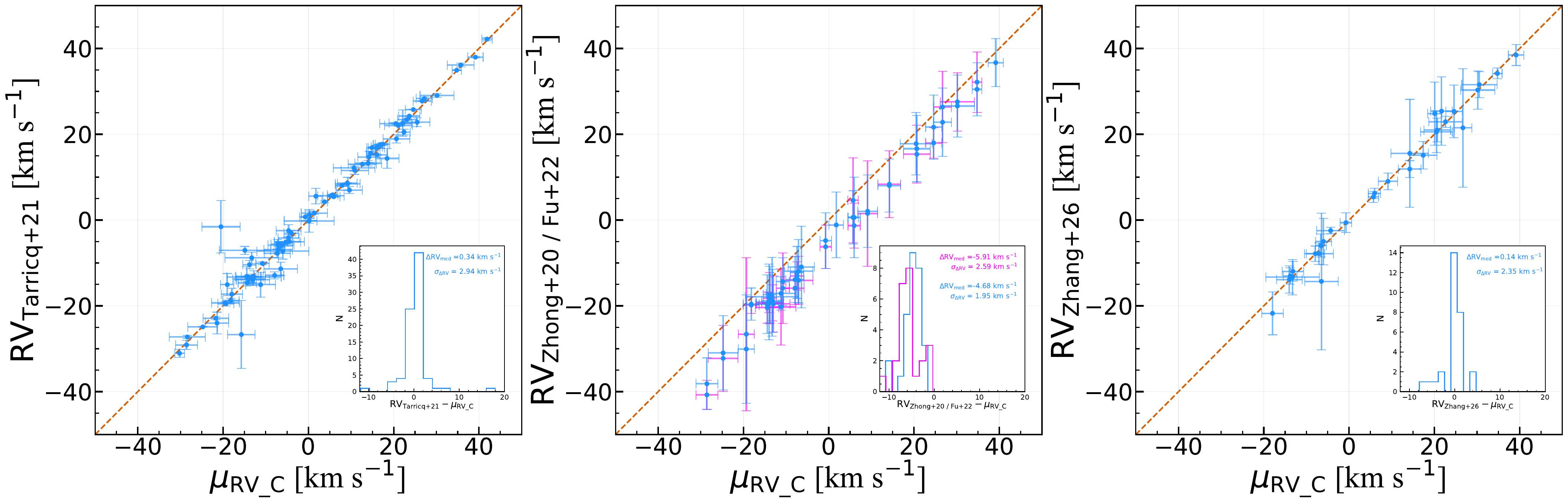}
\caption{Comparison of our cluster average RVs with literature values. The left, middle, and right panels show the comparisons with \citet{2021A&A...647A..19T}, both \citet{2020A&A...640A.127Z} (purple dots) and \citet{2022A&A...668A...4F} (blue dots), and \citet{2026RAA....26e5001Z}, respectively. Distributions of the RV differences are shown as inset histograms in each panel. The red dashed lines are 1:1 scale lines.}
\label{Fig6}
\end{figure*}

To ensure data reliability, we excluded clusters for which the literature RV was computed from fewer than three stars. We performed a cross-match between our sample and the four aforementioned catalogs to identify the shared OCs, yielding 82, 24, 28, and 29 common clusters, respectively. Fig.~\ref{Fig6} shows the distribution of differences between the average RVs derived from our color-filtered member star and those reported in the four catalogs. For the comparison with \citet{2021A&A...647A..19T} (left panel), the median RV difference is 0.34\,km\,s$^{-1}$ with a scatter of 2.94\,km\,s$^{-1}$, indicating that our RV results are consistent with theirs. For the comparisons with \citet{2020A&A...640A.127Z} and \citet{2022A&A...668A...4F} (middle panel), the median RV differences are -5.91 and -4.68\,km\,s$^{-1}$, with scatters of 2.59 and 1.95\,km\,s$^{-1}$, respectively. The large RV offsets primarily reflect an $\sim$5\,km\,s$^{-1}$ systematic difference between the LAMOST LRS and {\it Gaia} RV zero points \citep{2022A&A...659A..95T}. In contrast, the comparison with \citet{2026RAA....26e5001Z} (right panel), which relies on LAMOST MRS, yields a median RV offset of merely 0.14\,km\,s$^{-1}$ with a scatter of 2.35\,km\,s$^{-1}$. This confirms the absence of a significant zero-point offset between our derived RVs and the LAMOST MRS data.

Overall, the average RVs derived in this work demonstrate broad consistency with existing literature, particularly with those based on {\it Gaia} RVs. The noticeable offsets observed in certain comparisons primarily reflect well-documented systematic differences between survey instruments, such as the zero-point shift between LAMOST and {\it Gaia}, consistent with previous findings.

\subsection{The RV from binary stars}

It is well established that the measured RV of a binary system generally includes an additional component due to orbital motion. Previous studies (e.g., \citealt{2008AJ....135.2264G,2024AJ....167..212K}) have explicitly pointed out that the observed RVs of unresolved binaries can significantly inflate the cluster RV dispersion, which may lead to incorrect inferences about the system's dynamical state. To assess the impact of binary stars on the RV of OCs, we employed a similar method to that used in our previous work \citet{2025A&A...698A.295T} to exclude binaries with a high mass ratio (q$>$0.5). We used the PARSEC 1.2S isochrones \citep{2017ApJ...835...77M} to compute the binary sequence for each cluster. In Fig.~\ref{Fig1} as an example, the orange dots are identified as binary stars with a mass ratio q$>$0.5.

\begin{figure}[!htbp]
\centering
\includegraphics[width=0.95\textwidth]{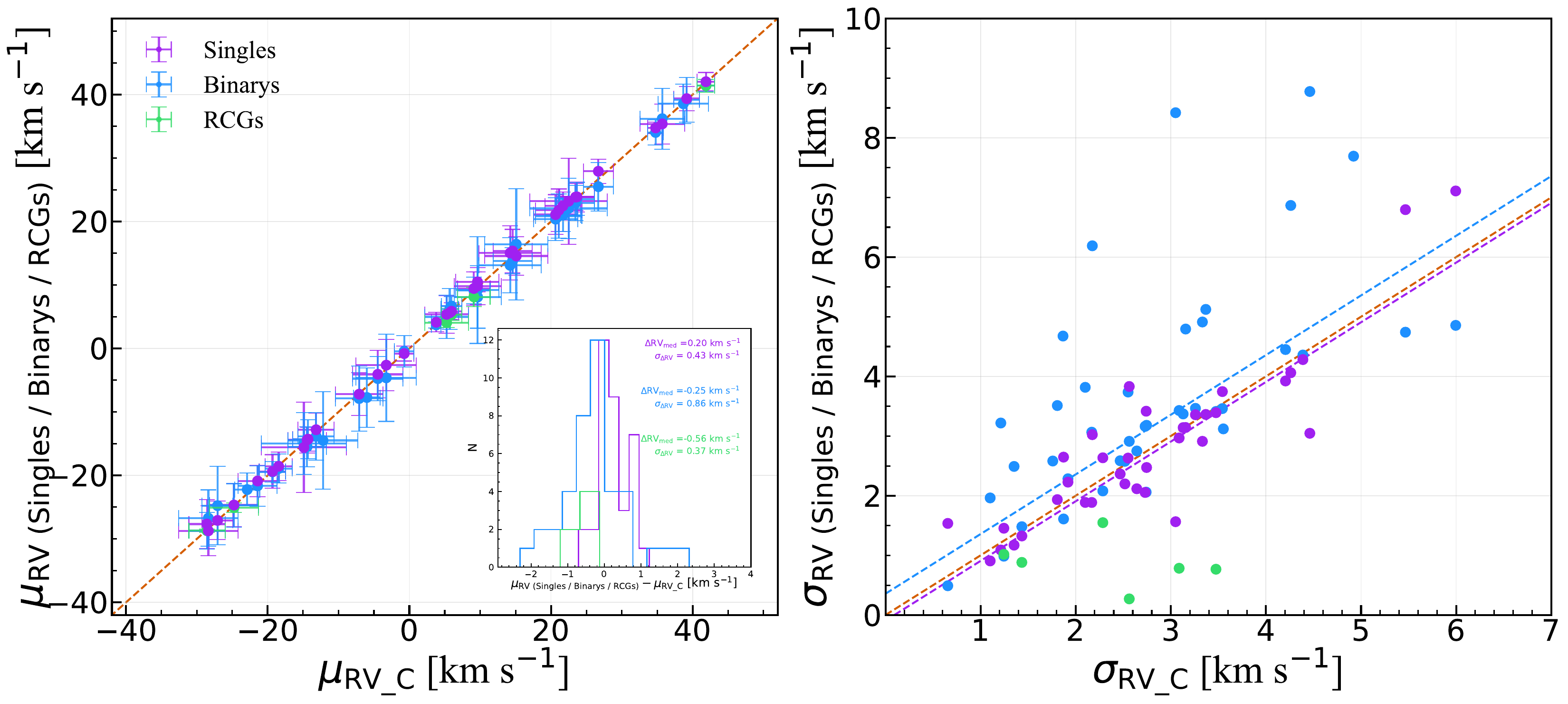}
\caption{Comparison of average RVs (left panel) and RV dispersion (right panel) derived from singles, binaries, and red clump giants with those derived from member stars. Purple and blue points represent the results from the color-filtered single stars and binary stars, respectively. The green points represent the results from the red clump sample. Distributions of the RV differences are shown as inset histograms in the left panel. The red dashed lines are 1:1 scale lines, while the blue and purple dashed line indicates the same relation shifted upward by the offset of 0.4\,km\,s$^{-1}$ and -0.1\,km\,s$^{-1}$}
\label{Fig7}
\end{figure}

To investigate the impact of binaries on the average RVs and RV dispersions of OCs, we applied the same color-based filtering to the binary member stars. We retained only clusters containing more than 20 binary members with RV measurements within the intrinsic color range of [0.2, 1.2]\,mag, resulting in a final sample of 40 clusters. The blue dots in Fig.~\ref{Fig7} illustrate the average RVs and RV dispersions derived from the binary subsample, plotted against the corresponding estimates from the full color-filtered sample. As shown in the left panel of Fig.~\ref{Fig7}, there is a median RV offset of -0.23\,km\,s$^{-1}$ with a scatter of 0.86\,km\,s$^{-1}$. While the two samples yield highly consistent average RVs for the OCs, the binary subsample exhibits a systematically larger RV dispersion, which aligns with theoretical expectations (right panel of Fig.~\ref{Fig7}). This suggests that the effect of the RV of the binary orbital motion on the cluster's average RV is essentially negligible. However, their contribution becomes non-negligible when considering the OC velocity dispersion.

As a comparison, we similarly performed Gaussian fitting on the member stars after excluding the high mass ratio binaries, applying a 3$\sigma$ clipping criterion and requiring a minimum of 20 retained members, which ultimately left 36 clusters in the sample, as indicated by the purple points in Fig.~\ref{Fig7}. It should be clarified that although the label ``Singles'' is used in the figure, these samples are not strictly pure. Due to the limitations of the photometric filtering method, we are unable to exclude low mass ratio binaries, and hence the sample remains contaminated by them. Overall, the median offset of the average RV obtained from the single star sample is 0.2\,km\,s$^{-1}$, with a corresponding dispersion of 0.43\,km\,s$^{-1}$, as shown by the purple points in the left panel of Fig.~\ref{Fig7}. Meanwhile, the RV dispersion shows a slight decrease, with a median value of -0.1\,km\,s$^{-1}$, as indicated by the purple dashed line in the right panel of Fig.~\ref{Fig7}.

It should be noted that, limited by the method we used, the binary sample employed for the calculation only includes high mass ratio binaries. Low mass ratio binaries are difficult to identify; therefore, the actual impact of binary orbital motion on the measured radial velocity dispersion might be underestimated, introducing a bias into the current results. We look forward to the upcoming {\it Gaia} DR4, as its all-epoch RV measurements may be used to identify the low mass ratio binaries. This future dataset will enable us to more accurately investigate the true impact of unresolved binaries on the dynamical properties of open clusters.

\subsection{The RV from red clump giants}\label{sec:RVRCG}

In addition to the impact of binaries on OCs' RV dispersions, members in different evolutionary phases can be exploited to trace the cluster systemic RV and the corresponding dispersion. Red clump giants define a compact and readily identifiable locus in open cluster CMDs, and thus provide a clean tracer of the cluster kinematics \citep{2016MNRAS.458.3150C,2022A&A...658A..14C}. Their high luminosities generally ensure high-S/N {\it Gaia} RVS observations in the Ca\,\rm\uppercase\expandafter{\romannumeral2} triplet, yielding particularly precise RV measurements \citep{2023A&A...674A...5K,2023A&A...674A...6S}. We selected OCs with red clump members and estimated the cluster's average RVs and RV dispersions using those red clump giants. In total, we selected six OCs: OCSN 207 (IC 4756), OCSN 221 (NGC 752), OCSN 236 (NGC 3532), OCSN 241 (NGC 6633), OCSN 261 (Ruprecht 147), and OCSN 265 (Stock 2). {\it Gaia} DR3 provides RV measurements for all 59 red clump giants, with uniformly high precision; the median RV uncertainty is e$\_{\rm RV}\approx 0.15$\,km\,s$^{-1}$. The results for these six clusters are shown in Table.~\ref{table2}.

\begin{deluxetable*}{crrrrrrc}
\tabletypesize{\normalsize}
\renewcommand{\arraystretch}{1.1}
\tablecaption{The average RVs and RV dispersions of the six OCs derived from red clump giants. \label{table2}}
\tablehead{
\colhead{Name} & \colhead{RAdeg} & \colhead{DEdeg} & \colhead{$\sigma_{\rm R.A.}$} & \colhead{$\sigma_{\rm Decl.}$} & \colhead{$\mu_{\rm RV\_RCG}$} & \colhead{$\sigma_{\rm RV\_RCG}$} & \colhead{N$_{\rm RV\_RCG}$} \\
\colhead{} & \colhead{[deg]} & \colhead{[deg]} & \colhead{[km\,s$^{-1}$]} & \colhead{[km\,s$^{-1}$]} & \colhead{[km\,s$^{-1}$]} & \colhead{[km\,s$^{-1}$]} & \colhead{} \\
\colhead{(1)} & \colhead{(2)} & \colhead{(3)} & \colhead{(4)} & \colhead{(5)} & \colhead{(6)} & \colhead{(7)} & \colhead{(8)}
}
\startdata
OCSN 207 & 279.67 &   5.43 & 0.5  & 0.77 & -25.08 & 0.77 & 15  \\
OCSN 221 &  29.16 &  37.78 & 0.67 & 0.78 &   5.32 & 0.88 & 13  \\
OCSN 236 & 166.39 & -58.69 & 1.03 & 0.96 &   4.06 & 0.78 &  7  \\
OCSN 241 & 276.85 &   6.65 & 0.63 & 0.58 & -28.7  & 0.27 &  6  \\
OCSN 261 & 289.07 & -16.42 & 0.37 & 0.79 &  41.4  & 1.02 &  9  \\
OCSN 265 &  33.88 &  59.52 & 1.01 & 1.02 &   8.12 & 1.55 &  9  \\
\enddata
\end{deluxetable*}

As shown by the green dots in Fig.~\ref{Fig5}, we compare the RV dispersion derived from red clump giants with the tangential velocity dispersion. The results indicate that the RV dispersion estimated from red clump giants is comparable to the tangential velocity dispersion. The green dots in Fig.~\ref{Fig7} show a comparison between the average RVs and RV dispersions derived from red clump giants and the results obtained from the color-based filtering of member stars. The RV differences have a median of -0.56\,km\,s$^{-1}$ and a scatter of 0.37\,km\,s$^{-1}$. These results provide additional evidence that the RVs derived from red clump giants are not only consistent with those obtained from member stars, but that the RV dispersion is much smaller than the dispersion characterized by the member stars. 

To explore the cause of this phenomenon in more detail, we cross-matched the RVS spectra for the 59 red clump giants; however, {\it Gaia} DR3 has released spectral data for only six stars. We will present the mean RVS spectra of these six stars in the Appendix~\ref{Appendix}. We find that the spectra of these red clump stars are all of high quality, with a prominent Ca\,\rm\uppercase\expandafter{\romannumeral2} triplet. These red clump giants are predominantly early K-type stars, for which the Ca\,\rm\uppercase\expandafter{\romannumeral2} triplet is relatively prominent. In addition, their magnitudes are sufficiently bright to ensure high quality RVS spectra, resulting in more accurate RV measurements. Given the high quality RVS spectra of OCs' red clump giants, using the red clump giants sample to characterize the average RV and RV dispersion of OCs is in line with expectations.

\section{Conclusions}\label{sec:C}

Although {\it Gaia} DR3 provides RVs for more than 33 million stars with relatively high precision, the Ca\,\rm\uppercase\expandafter{\romannumeral2} triplet in both hot and cool stars can be blended with other spectral lines, potentially biasing the measured RVs \citep{2018A&A...616A...5C,2023A&A...674A...5K,2023A&A...674A...7B}. In OCs, this bias manifests as a markedly overestimated RV dispersion (see the right panel of Fig.~\ref{Fig3}). To mitigate the impact of such biases, we re-examine the average RVs and RV dispersions of OCs within 500\,pc of the solar neighborhood by applying a color-based filter to the member stars. When compared against cluster average RVs derived from high-resolution spectroscopic data, our results are broadly consistent with literature values, exhibiting a median offset of only 0.34\,km\,s$^{-1}$ relative to \citet{2021A&A...647A..19T}. In contrast, when compared with cluster average RVs based on low-resolution LAMOST spectra \citep{2020A&A...640A.127Z,2022A&A...668A...4F}, our average RVs show a systematic offset of about 5\,km\,s$^{-1}$, as shown in Fig.~\ref{Fig6}. Additionally, our results show no zero-point offset compared to those from the LAMOST MRS \citep{2026RAA....26e5001Z}.

We further find that the color-based filter markedly reduces the RV dispersion. As shown in Fig.~\ref{Fig3}, for our sample the median dispersion decreases from 3.76\,km\,s$^{-1}$ prior to color-based filtering to 2.79\,km\,s$^{-1}$ afterward. The median value of the RV dispersion is reduced by 26\%. Although the method can effectively reduce the RV dispersion of OCs, it remains larger than the tangential velocity dispersion, as shown in Fig.~\ref{Fig5}. The main reasons may include residual biases in {\it Gaia}'s RV measurements within this color range, as well as the incomplete removal of binaries from our sample. In addition, the average RVs of OCs derived from binary stars are nearly identical to those from the member stars, whereas the binary sample exhibits a larger RV dispersion, as expected. It is difficult to eliminate the effect of binaries on the RV dispersion; however, we can look forward to the release of full-epoch RV measurements in {\it Gaia} DR4 to further account for the influence of binary stars.

Notably, for red clump giants, the Ca\,\rm\uppercase\expandafter{\romannumeral2} triplet is prominent, and these stars are sufficiently bright, resulting in very high-quality RVS spectra and high-precision RV measurements. The average RVs of OCs derived from red clump giants are not only consistent with those obtained from member stars, but also exhibit a significantly reduced RV dispersion of $\sim$1.6\,km\,s$^{-1}$, which is nearly identical to the tangential velocity dispersion, as indicated by the green dots in Fig.~\ref{Fig5}. We thus argue that utilizing {\it Gaia} RVs for red clump giants represents an effective Gaia-only strategy to trace both the average RVs and RV dispersions of their host clusters.

Therefore, although {\it Gaia} DR3 provides unprecedented RV data for OCs, robust estimates of average RV and RV dispersions of OC require careful selection of members. By applying a color-based filter, we obtain cluster average RVs consistent with other spectroscopic surveys and substantially reduce the inflated RV dispersions by about 26\%. Moreover, red clump giants, whose {\it Gaia} RV dispersions closely match the tangential dispersions, provide the most reliable tracers of both average RVs and RV dispersions.

\begin{acknowledgments}
Yu Zhang acknowledges the support from the science research grants from the Central Government Guidance Fund for Local Sci-Tech Development (No. ZYYD2025QY27), Chinese Academy of Sciences (CAS) "Light of West China" Program (No. 2022-XBQNXZ-013), and the Talent Team Program of Xinjiang Talent Development Fund.
J. Z. would like to acknowledge the Shanghai Oriental Talents Program(QNKJ2025055) and the science research grants from the China Manned Space Project with NO. CMS-CSST-2025-A19. 
Guimei Liu would like to acknowledge financial support from the China Scholarship Council (Grant Nos. 202504910181). This work has made use of data from the European Space Agency (ESA) mission {\it Gaia} (\url{https://www.cosmos.esa.int/gaia}), processed by the {\it Gaia} Data Processing and Analysis Consortium (DPAC, \url{https://www.cosmos.esa.int/web/gaia/dpac/consortium}). Funding for the DPAC has been provided by national institutions, in particular, the institutions participating in the {\it Gaia} Multilateral Agreement.
\end{acknowledgments}

\software{astropy \citep{2013A&A...558A..33A,2018AJ....156..123A,2022ApJ...935..167A}, numpy \citep{2020Natur.585..357H}, scipy \citep{2020NaMet..17..261V}, matplotlib \citep{4160265}}


\appendix

\section{The mean RVS spectra of red clump gaints}\label{Appendix}
\begin{figure*}[!htbp]
\centering
\includegraphics[width=0.99\textwidth]{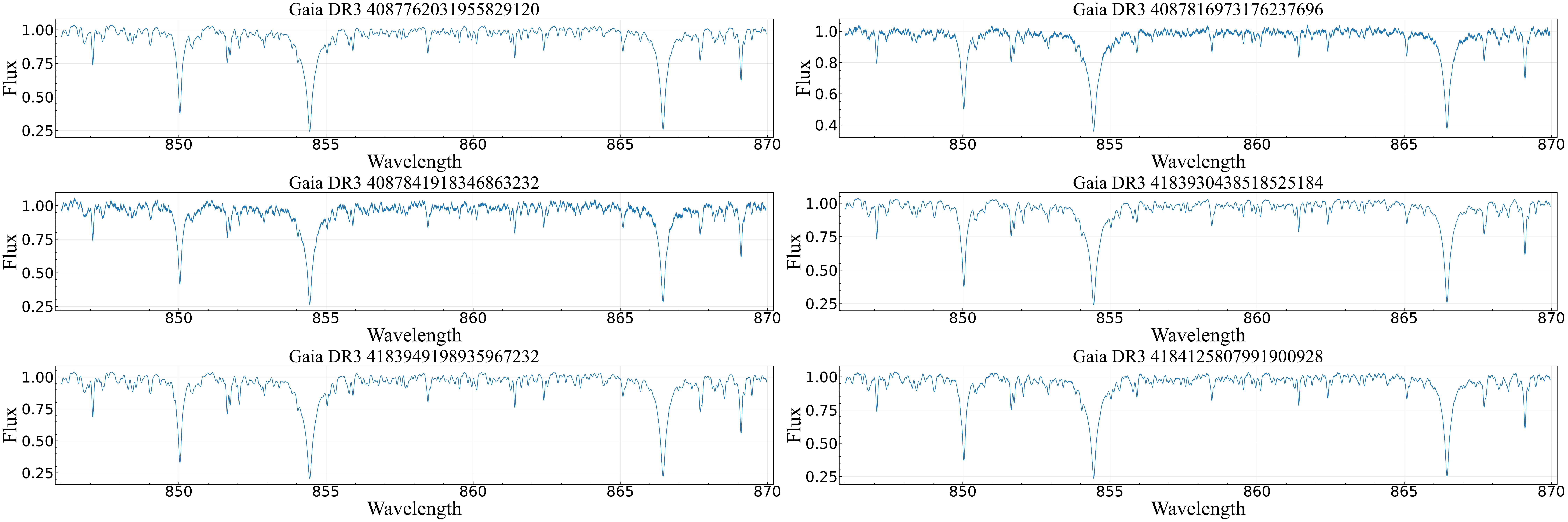}
\caption{The {\it Gaia} DR3 mean RVS spectra of the six red clump giants mentioned in Sect.~\ref {sec:RVRCG}.}
\label{FigA}
\end{figure*}

Based on the 59 red clump giant members identified in the six OCs considered here, we only downloaded the RVS spectra of 6 members from the {\it Gaia} Archive\footnote{\url{https://gea.esac.esa.int/archive/}} and plotted them in Fig.~\ref{FigA}.

\bibliography{PASPsample701}{}
\bibliographystyle{aasjournalv7}



\end{document}